# PROTEOTRONICS: ELECTRONIC DEVICES BASED ON PROTEINS

ELEONORA ALFINITO, LINO REGGIANI

*Università del Salento*

JEREMY POUSSET

*CNR-IMM (Lecce)*

The convergent interests of different scientific disciplines, from biochemistry to electronics, toward the investigation of protein electrical properties, has promoted the development of a novel bailiwick, the so called proteotronics. The main aim of proteotronics is to propose and achieve innovative electronic devices, based on the selective action of specific proteins. This paper gives a sketch of the fields of applications of proteotronics, by using as significant example the detection of a specific odorant molecule carried out by an olfactory receptor. The experiment is briefly reviewed and its theoretical interpretation given. Further experiments are envisioned and expected results discussed in the perspective of an experimental validation.

## 1. Introduction

Recently, science has changed its horizons, switching off overspecialization, and embracing a syncretic perspective. This movement sees the convergence of multiple interests, know-how and know-why, and therefore the need of a common language, with new key words. In particular, in the field of electronics, more and more pressing requirements push toward the devising and setting up of very sensitive and fast detectors, useful for noninvasive diagnostics in medicine, food security etc. Biointegrated micro/nano-devices are the way followed by several groups for exploiting this trend. Accordingly, the new branch of science, conjugating electronics and proteomics, the large-scale study of proteins, was given the name of *proteotronics* [1].

## 2. Focus on experiments

Recent investigations performed with chemical and physical approaches have definitely confirmed the possibility of detecting the protein activity by using electrical measurements [2,3]. As a matter of fact, in spite of a completely different topological structure (highly not periodic), proteins seem to exhibit a conductivity like medium-gap semiconductors [1,2], with electrical features deeply depending on protein conformation and environmental conditions.

Proteins show very specific 3D structures in their native states, which, in the presence of specific stimuli, can significantly modify (active states). The detection of this conformational change is monitored by different techniques, both chemical-like, as electrochemical impedance spectroscopy (EIS) [3], and physical-like, as current-voltage or conductance measurements [4,5]. In both the cases, the experimental set-up involves protein receptors anchored on specific substrates and used to detect the *in vitro* response to specific stimuli (small molecules capture or light absorption ). In the following we recall the main data concerning the dose-response of rat OR-I7, used as a preliminary test for the development of a bio-electronic nose [3]. OR-I7 is an olfactory receptor, highly specific for the aldehyde group, showing a very good response to octanal. Samples were prepared by immobilizing OR-I7 proteins, in their membrane fraction, on a gold electrode, by using a self-assembled multilayer. The measurements were performed in an electrochemical cell whose working electrode is constituted by the functionalized gold electrode.

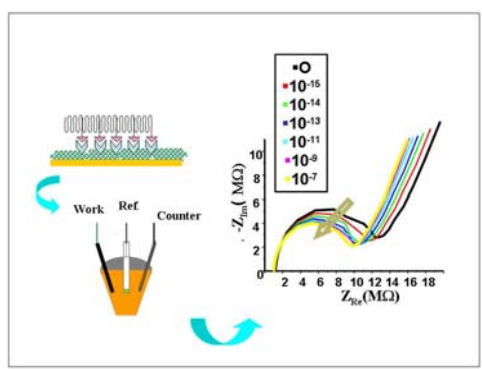

Figure 1. Schematic representation of EIS measurements performed on samples of rat OR -I7 at increasing concentrations of the specific odorant, octanal. On the left: the sample and the electrochemical cell; on the right: the experimental results. The straight arrow indicates the increasing concentration of octanal.

The sensitivity to octanal has been explored over a range of about 9 orders of magnitude of odorant concentration. The main results are a good selectivity with respect to other aldehydes, and a monotonic increase of sensitivity at increasing concentration. Figure 1 reports the experimental Nyquist plots at increasing octanal concentration (expressed in Molar). Data are analyzed by using a macroscopic circuit analogue called Randles cell, where the element most sensitive to the concentration variation is the polarization resistance : it decreases for about 18% at an octanal concentration of $10^{-4}$ M .

### 3. Focus on theory

A theoretical framework able to include all known results and predict new outcomes is the web on which proteotronics can develop. Accordingly, an numerical approach able to model the main electrical mechanisms revealed by experiments on proteins *in vitro* was set up [1,2]. It focuses on the interactions between amino acids, that constitute the protein building-blocks, and as such are responsible of the specific sensing action of the given protein.. As matter of fact, it is sufficient to change a single amino acid for producing a ill-functioning objects, like it happens for defective hemoglobins. These interactions may be described like a network covering the whole protein . In the case of interactions of electrical origin the simplest way to describe this web is the impedance network protein analogue (INPA) [1,2] which, analogously to the well known Hodgkin-Huxley model, describes the interactions by using a set of linear equations, solved by a computational procedure. This procedure is based on the Kirchhoff's laws and therefore the equations are linear. Elements of non-linearity may be introduced by using stochastic terms within a Monte Carlo solving approach. More details of this approach can be found in Ref. [2]. Below the INPA is applied to the case of rat OR-I7. The model needs a few information concerning the protein under examination: the 3D structure, the kind of interaction between amino acids, the resistivity and polarizability of each amino acid. The model contains a free parameter, the interaction radius $R_C$, whose value can be correlated to the ligand dose [1,2]. By using these input data it is possible i) to reproduce experimental results with a satisfactory agreement; ii) to predict the electrical properties of novel structures. In doing so, $R_C$ is selected by the comparison of calculated data with experiments and its value is used to foresee new responses. In particular, the experimental results shown in previous section, select the value $R_C=32$ Å, as reported in the inset of Fig. 2. This figure shows the calculated Nyquist plot of rat OR I7 in the native and active state. The calculated real impedance (ReZ) corresponds to the measured polarization resistance. The same figure reports the I-V characteristics of rat OR I7 calculated with $R_C=32$ Å, for the native and active state, to simulate the response in the presence/absence of octanal. The I-V characteristics are super-linear and this is attributed to a tunneling mechanism of charge transport [1,2,4]. In particular, by using a wide range of bias values, it is possible to observe two different tunneling regimes, the direct tunneling (DT) and the Fowler-Nordheim (FN). An efficient strategy for an accurate investigation of the DT-FN transition is given by the analysis of conductance fluctuations. This analysis is quite useful both for the general aim of a most deep understanding of the internal mechanisms driving the current transport [1] but also for an early monitoring of the protein response under a bias stress. Figure 3 reports the conductance fluctuation distribution (CFD) over a range of 20 orders of magnitude, clearly showing the DT regime, below 0.1 V, and the FN regime (in the inset) above 0.5 V. The change of regime is very sharp: in the low conductance region CDF has the same characteristic shape for more than 17 magnitude orders, a protein fingerprint. At higher conductance values the competition between DT and FN is indicated by the presence of a secondary peak. Finally, a new linear regime is established and a single sharp distribution appears.

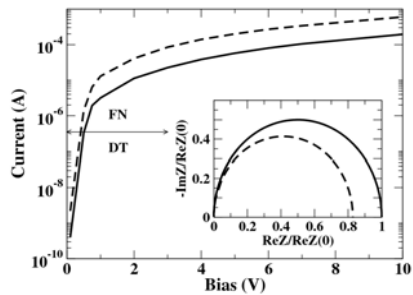 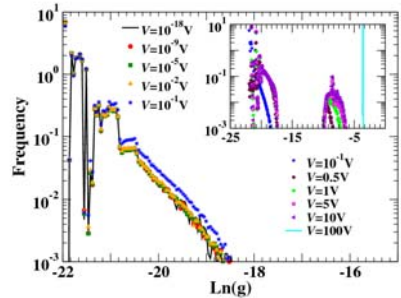

Figure 2. I-V characteristics for rat OR I7 in the native (continuous line) and active (dashed line) states. In the inset: Nyquist plots for the same templates.

Figure 3. Distribution of conductance fluctuations at different bias values for rat OR I7 in the native state.

## 4. Conclusions

Proteotronics is here introduced as a new branch of electronics devoted to investigate the electrical properties of proteins with the aim to develop new bio-electronic devices. As significant application, the case of rat olfactory receptor OR I7 acting as an odor sensor is considered. From the theoretical side, proteotronics has been used to predict the OR-I7 current-voltage characteristics in a wide range of applied voltages. The predicted I-V superlinear characteristics and the extreme conductance fluctuations still wait to be confirmed from an experimental side.

## Acknowledgments

This research is supported by the EC under the Bioelectronic Olfactory Neuron Device (BOND) project within the grant agreement number 228685-2.